\documentclass[epj]{svjour}
\usepackage{latexsym}
\usepackage{graphics}
\usepackage[]{graphicx} 
\usepackage{mathrsfs}
\usepackage{amsmath}
\begin{document}
\title{Calculated fission-fragment mass yields and average total kinetic energies of
  heavy and superheavy nuclei}
\author{Martin Albertsson \inst{1}\textsuperscript{,}\thanks{\email{martin.albertsson@matfys.lth.se}}, B. Gillis Carlsson\inst{1},
  Thomas D{\o}ssing\inst{2}, Peter M\"{o}ller\inst{1}, J{\o}rgen Randrup\inst{3}, \and Sven {\AA}berg\inst{1}
}                     
%
%
\institute{Mathematical Physics, Lund University, S-221 00 Lund, Sweden \and
  Niels Bohr Institute, University of Copenhagen, DK-2100 Copenhagen {\O}, Denmark \and
  Nuclear Science Division, Lawrence Berkeley National Laboratory, Berkeley, California 94720, USA}

\date{Received: date / Revised version: date}
%

\abstract{Fission-fragment mass and total-kinetic-energy (TKE) distributions
following fission of even-even nuclides in the region $74 \leq Z \leq 126$ 
and $92 \leq N \leq 230$, comprising 896 nuclides have been
calculated using the Brownian shape-motion method. 
The emphasis is the region of superheavy nuclei. 
To show compatibility with
earlier results the calculations are extended to include earlier studied regions. 
An island of asymmetric fission is obtained in the superheavy
region, $106\leq Z\leq114$ and $162\leq N\leq 176$,
where the heavy fragment is found to be close to $^{208}$Pb and the light fragment adjusts accordingly.
Most experimentally observed
$\alpha$-decay chains of superheavy nuclei with $Z > 113 $ terminate by spontaneous
fission in our predicted region of asymmetric fission. 
In these cases, the pronounced large asymmetry is accompanied by a low TKE value compatible with measurements.
}

\authorrunning{ }
\titlerunning{ }
\maketitle
\section{Introduction}
\label{intro}
The seven new elements in the range $107 \le Z \le 113$ were all identified through
$\alpha$-decay chains ending in previously observed $\alpha$ decays.
However, for the still heavier elements created in $^{48}$Ca-induced fusion-evaporation reactions
it is more involved to establish the specific isotope created since
most $\alpha$-decay chains end in spontaneous fission  and not in
a previously known $\alpha$ decay. 
For an overview, see refs.\ \cite{dullmann15:a,rudolph16:a}.
To contribute to the interpretation of these experimental results
we calculate 
fission-fragment mass and kinetic-energy distributions
and average total kinetic-energies (TKE) for 896 even-even nuclides 
in the region $74 \leq Z \leq 126$ and $92 \leq N \leq 230$.
We use the Brownian shape-motion (BSM) method 
\cite{randrup11:a} which has been extensively compared to experimental data \cite{randrup13:a}
and used in studies of the ``new region of asymmetry'' in the neutron-deficient Pb region \cite{andreyev13:a,moller15:b}.

\section{Calculational details}
The first step is to calculate the potential-energy surfaces as functions
of five shape parameters. 
In previous publications 
\cite{randrup11:a,randrup13:a,moller15:b,randrup11:b,moller12:a,ward17:a}
potential-energy surfaces calculated in
ref.\ \cite{moller09:a} have been used. The current calculations are also done as specified in
ref.\ \cite{moller09:a} but with
one difference:
In the Strutinsky shell-correction procedure we use a larger smoothing range
\begin{equation}
   \gamma = 1.5\times\frac{41\text{ MeV}}{A^{1/3}}B_{\rm S},
  \label{smooth}
\end{equation}
where $B_{\rm S}$ is the ratio of the surface area of the current shape to
that of a spherical shape.
This is particularly important for nuclei in the vicinity of
fermium (Fm) as discussed in great detail in refs.\ \cite{moller89:c,moller94:b}.

In the calculation, most of the CPU time is used to calculate
the single-particle levels. Once the levels are determined, the
time needed to calculate the shell and pairing corrections and macroscopic
contributions to the potential energy is almost negligible.
Therefore we use the same set of levels to calculate the shell
corrections for several neighbouring nuclei, which still leads  to satisfactory
accuracy. We use levels calculated for four ``center'' nuclei, namely $^{214}$Rn, $^{288}$Pu,
$^{258}$Fm, and $^{270}$Hs. 
This strategy is routinely employed when calculating the potential energy for neighbouring nuclei \cite{randrup11:a,moller09:a,nilsson69:a,bolsterli72:a,moller89:a,moller01:a} 
and is sufficiently accurate for our overview here. As pointed out in sect.\ \ref{sect:res_fragmass},
we obtain very similar results as ref.\ \cite{moller15:b}, for the region
where our two studies overlap, even though potentials for
somewhat different nuclei are used to calculate the single-particle
levels and the corresponding shell corrections.

The critical neck radius, where fragment separation is assumed to occur,
is $c_0=1.5$ fm. To ensure that the vast majority of the random walks reach this neck radius before reaching 
the boundary of the employed shape lattice we have extended the grid with eight additional points in 
the elongation direction, corresponding to more than one million additional shapes.
The largest value of the dimensionless elongation parameter $q_2$ (defined in eq.\ (11) in ref.\ \cite{ward17:a})
is extended from roughly 16 to 20.

\subsection{Fission-fragment mass distributions} 
We calculate the fission-fragment mass distributions with
the BSM method on the multi-dimensional
potential-energy surfaces as specified in refs.\ \cite{randrup11:a,randrup13:a}.
The walks are started at the second minimum, when such a minimum
exists, otherwise at the ground-state minimum.
For the cases
where the ground-state shape is spherical, the walk
is started at the least deformed symmetric shape included in the grid,
corresponding to $\beta_2 \approx 0.12$.
We follow the prescription of the earlier study \cite{moller15:b} and
use the effective level density of ref.\ \cite{randrup13:a}.
A more refined calculation would employ
shape-dependent microscopic level densities as pioneered in ref.\ \cite{ward17:a} but
such a huge effort is not computationally feasible yet.
Also, we limit the study to even-even nuclei because yields
vary insignificantly between neighbouring nuclides except in a few isolated cases.
Each distribution is based on 10000 walks. Excitation energies are chosen
just sufficiently above the barrier to obtain reasonable
computing times. As in ref.\ \cite{moller15:b} the bias potential is 60 MeV.
The fission-fragment mass yield $Y(A)$ is defined as the percentage of events
resulting in fragment-mass number $A$.
The yield is normalized to 200\% because each fission event results in two fragments.

\subsection{Total kinetic energies \label{tkemeth}}
The total available energy in the fission process
is given by the initial excitation energy, $E_{\rm exc}$, with respect to the ground state of the fissioning nucleus and the $Q$ value,
\begin{equation}
\label{eq:qval}
Q^\ast=E_{\rm exc}+M(Z,N)-M(Z_{\rm L},N_{\rm L})-M(Z_{\rm H},N_{\rm H}),
\end{equation}
where $M(Z,N)$ is the ground-state mass of the parent nucleus,
and $M(Z_{\rm L},N_{\rm L})$ and $M(Z_{\rm H},N_{\rm H})$ are the ground-state masses of the
light and heavy fragments, respectively. 
The proton and neutron numbers, $Z$ and $N$, are determined by requiring the same $Z/N$ ratio as for the fissioning nucleus. 
In the present study only fragments with even $Z$ and $N$ are considered. 

The available energy is divided between the total kinetic energy and the total excitation
energy of the fragments, {\it i.e.},
\begin{equation}
\label{eq:qval2}
Q^\ast=E_{\rm TKE} + E_{\rm TXE}.
\end{equation}
The total excitation energy, $E_{\rm TXE}$, shared between the two fragments, is composed by the two parts,
intrinsic excitation energy and deformation energy of the light (L) and heavy (H) fragment, 
\begin{equation}
\label{eq:txe}
E_{\rm TXE}=E^*_{\rm sc} + E^{\rm def}_{\rm L}+E^{\rm def}_{\rm H}.
\end{equation}
The intrinsic excitation energy, $E^*_{\rm sc}$, is the energy difference between the total energy, $E_{\rm tot}$, 
and the potential energy at the scission configuration $U(\boldsymbol{\chi}_{\rm sc})$,
\begin{equation}
\label{eq:Esciss}
E^\ast_{\rm sc} =E_{\rm tot}-U(\boldsymbol{\chi}_{\rm sc}),
\end{equation}
where $\boldsymbol{\chi}$ denotes the five shape parameters.
The deformation energy is released in the fragments after scission when
the accelerated fission fragments relax their respective shapes to ground-state
deformations. The released energy is thereby added to the internal
excitation energy.

\begin{figure}[t] 
 \begin{center} 
 \includegraphics[width=\linewidth]{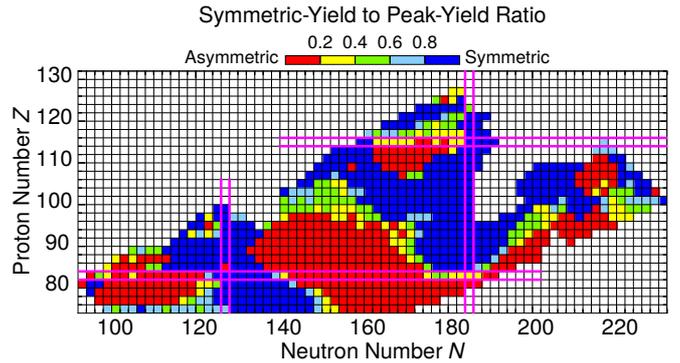} 
 \caption{Calculated symmetric-yield to peak-yield ratios versus $N$ and $Z$
   for fissioning nuclides between the proton and neutron
   drip lines and $74 \le Z \le 126$, for even-even nuclides. Nuclides with barriers
   calculated to be lower than 3 MeV are not included.
   Pairs of magenta parallel lines indicate magic neutron and proton numbers in the model 
   ($N=126,184$ and $Z=82,114$).
   } 
\label{s2pr} 
 \end{center} 
\end{figure}
\begin{figure}[b] 
 \begin{center} 
 \includegraphics[width=\linewidth]{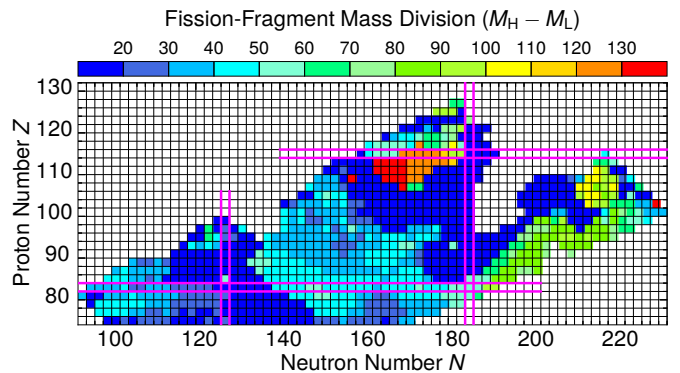} 
 \caption{Calculated heavy and light fission fragment mass differences
   $M_{\rm H} - M_{\rm L}$, following fission of heavy nuclei, analogous to fig.\ \ref{s2pr}
   in other respects.} 
\label{massdif} 
 \end{center} 
\end{figure}

The deformation energy of each fragment is calculated as the energy difference between the
fragment mass at scission \cite{albertsson19:c} and its ground-state mass, 
\begin{equation}
E^{\rm def}_i=M_i(\varepsilon^{\rm sc}_i)- M_i(\varepsilon^{\rm gs}_i),
\label{eq:deformation}
\end{equation}
where $i= {\rm L}$ or H. 

The masses in eqs.\ (\ref{eq:qval}) and (\ref{eq:deformation}), are calculated in the 
same macroscopic-microscopic model that was used to obtain the 
potential-energy surfaces \cite{moller04:a}. The fragment shapes at scission 
are taken as the spheroidal shapes characterized by the $\varepsilon_2$ values in the 5D shape parametrization, 
while the ground-state shapes include $\varepsilon_2, \varepsilon_4, \varepsilon_6$.

The $Q^*$ value is calculated from eq.\ (\ref{eq:qval}) and $E_{\rm TXE}$ from eq.\ (\ref{eq:txe}). 
The TKE value is then obtained from eq.\ (\ref{eq:qval2}).
The TKE yield $Y(E_{\rm TKE})$ is the percentage of fisson events per MeV kinetic energy,
while the average TKE is the kinetic energy of the relative motion of two fission fragments averaged over all fission events.

\begin{figure}[t] 
 \begin{center} 
 \includegraphics[width=\linewidth]{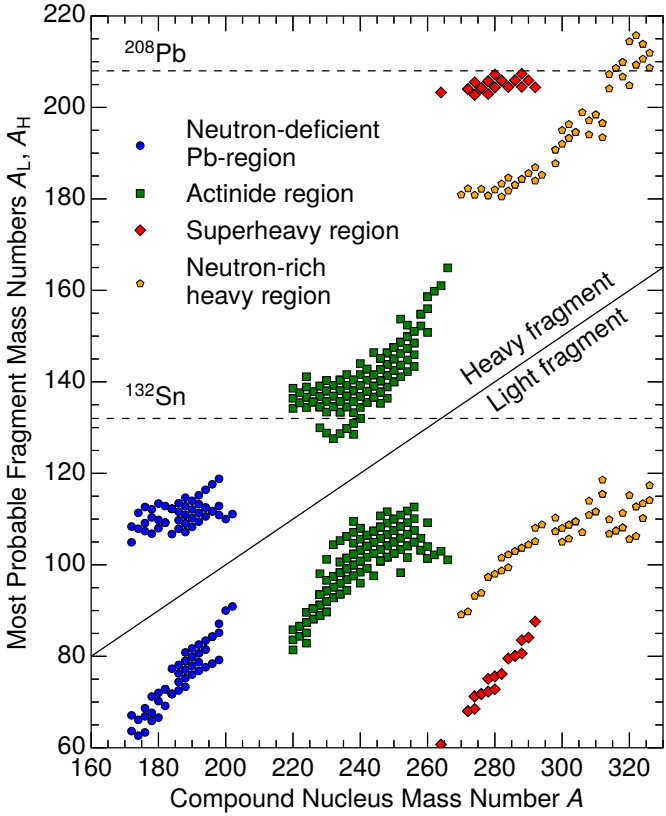} 
 \caption{
Calculated most probable fission-fragment mass numbers following fission of neutron-deficient($74\leq Z\leq 86$, $92\leq N\leq 126$), actinide ($74\leq Z\leq 96$, $132\leq N\leq 186$),
 superheavy ($106\leq Z\leq 114$, $156\leq N\leq 178$) and neutron-rich ($82\leq Z\leq 110$, $188\leq N\leq 218$) nuclides. 
   Only nuclides with asymmetric fission and
   with a symmetric-yield to peak-yield ratio less than 0.2
 are included (red squares in fig.\ \ref{s2pr}).}
\label{fragmasses} 
 \end{center} 
\end{figure}
\begin{figure}[b] 
 \begin{center} 
 \includegraphics[width=\linewidth]{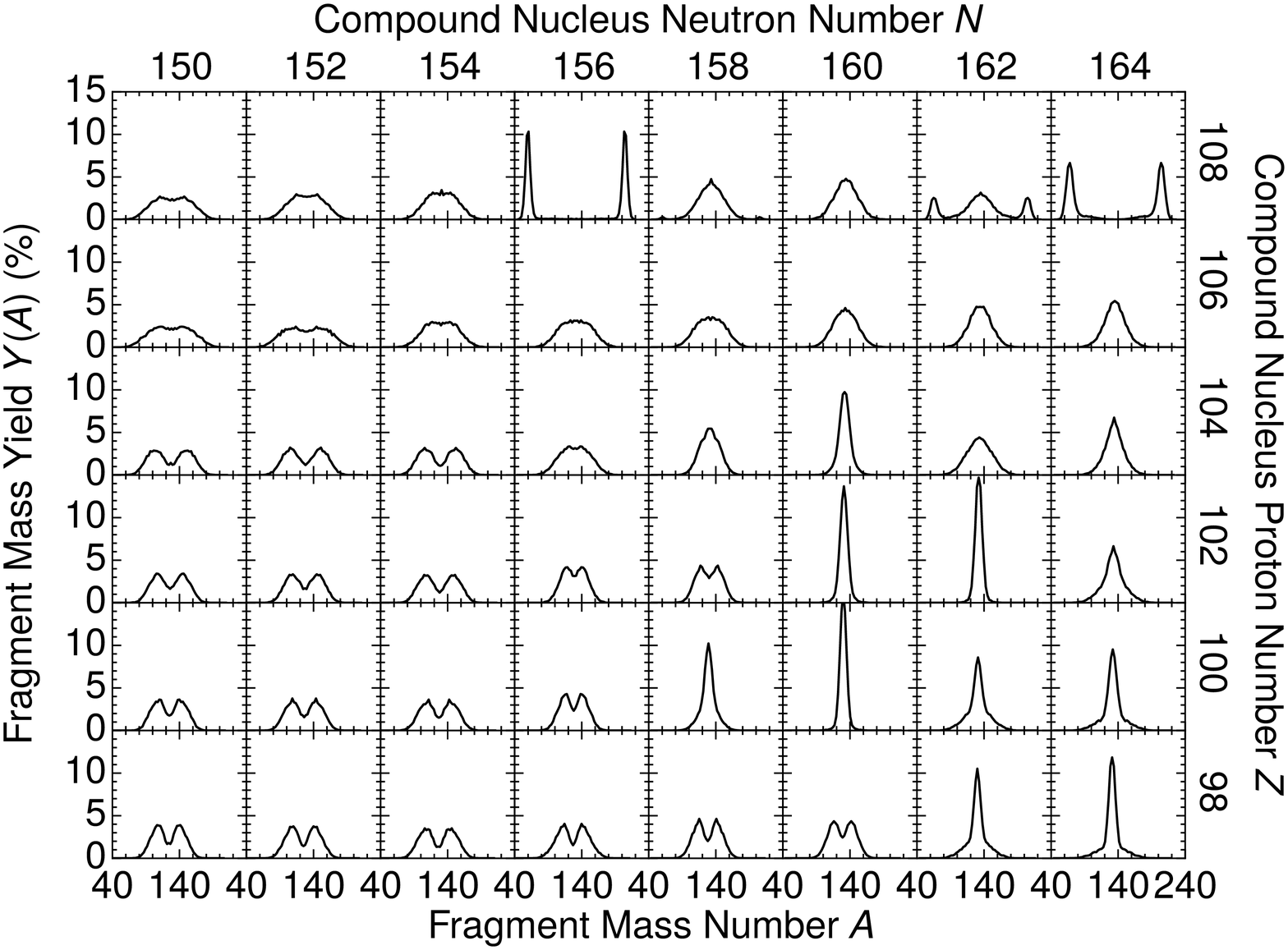} 
 \caption{Calculated fission-fragment mass distributions for fissioning
 nuclei in the region $98 \le Z \le 108$ and $150 \le N \le 164$.} 
\label{myields1} 
 \end{center} 
\end{figure}
\begin{figure}[t] 
 \begin{center} 
 \includegraphics[width=\linewidth]{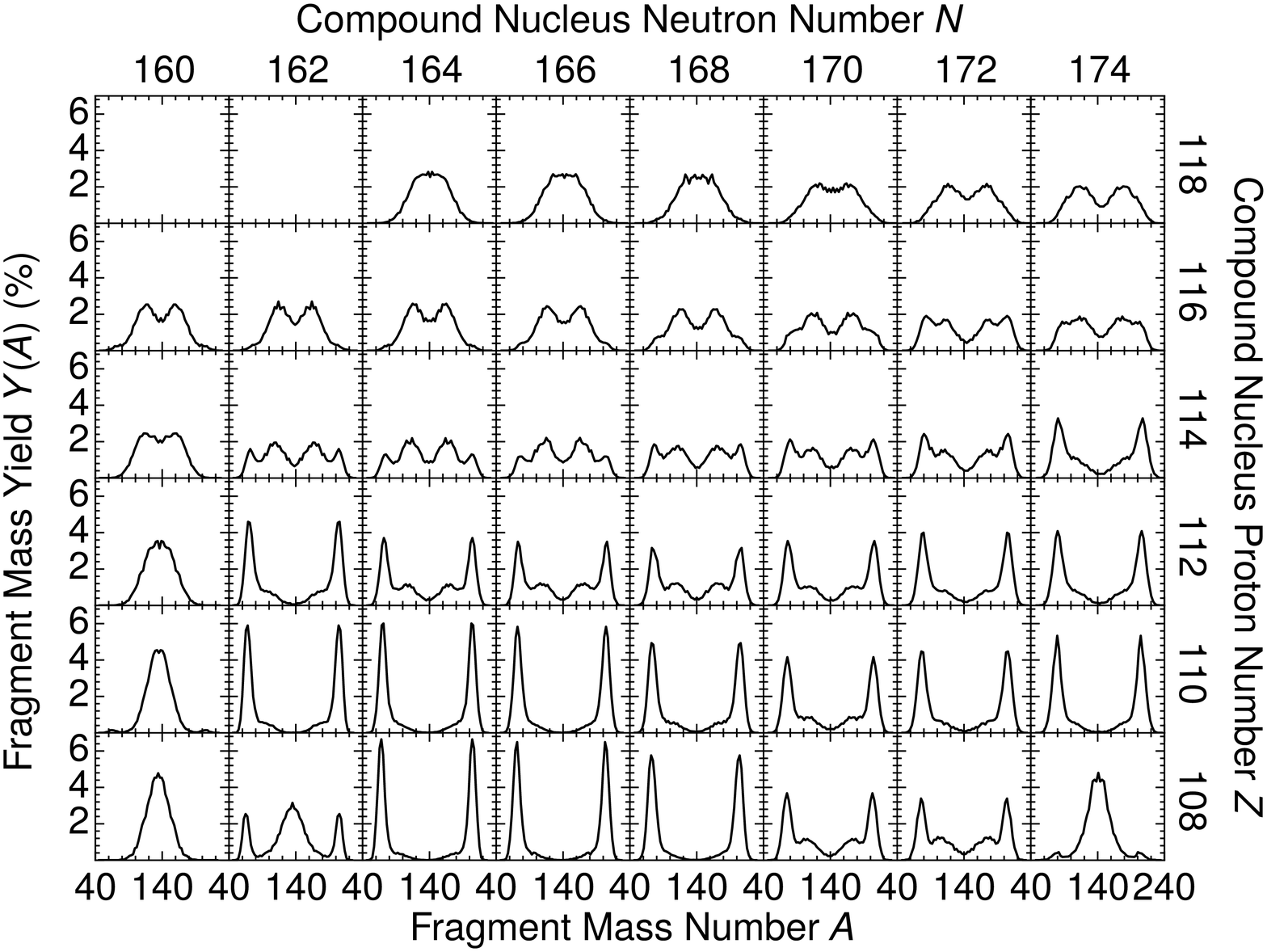} 
 \caption{Calculated fission-fragment mass distributions for fissioning
 nuclei in the region  $108 \le Z \le 118$ and $160 \le N \le 174$.} 
\label{myields2} 
 \end{center} 
\end{figure}
\begin{figure}[b] 
 \begin{center} 
 \includegraphics[width=\linewidth]{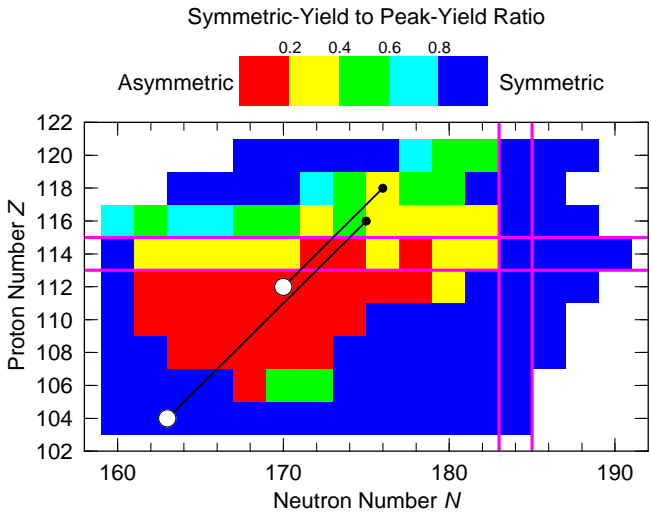}  
 \caption{Calculated symmetric-yield to peak-yield ratios versus $N$ and $Z$
   for nuclides in the region of observed superheavy elements, for even-even nuclides. 
   We show some representative observed decay chains with a small solid black dot
   indicating the start of the decay and the larger solid white dot indicating
   the termination by spontaneous fission.
   Pairs of thin  parallel lines indicate  magic neutron and proton numbers.} 
\label{s2prshe} 
 \end{center} 
\end{figure}
\begin{figure}[t] 
 \begin{center} 
 \includegraphics[width=\linewidth]{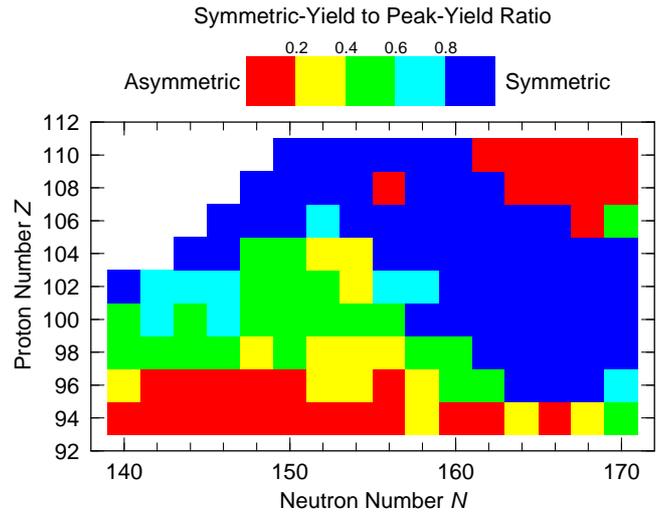} 
 \caption{Calculated symmetric-yield to peak-yield ratios versus $N$ and $Z$
   for even-even nuclides in the Fm region. } 
\label{s2prfm} 
 \end{center} 
\end{figure}
\begin{figure}[b] 
 \begin{center} 
 \includegraphics[width=\linewidth]{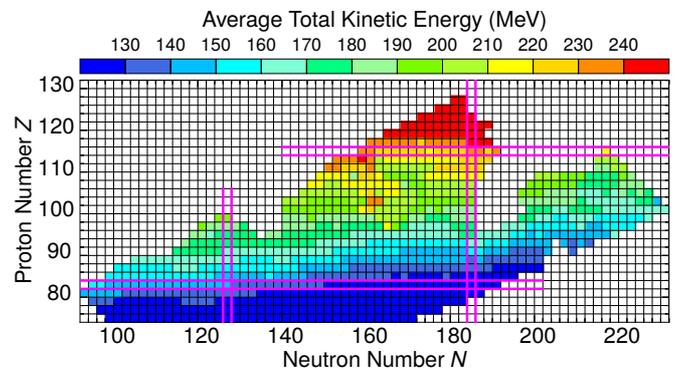} 
 \caption{Calculated average fission-fragment total kinetic energies following fission of
 heavy nuclei.}
\label{tkeav} 
 \end{center} 
\end{figure}
\begin{figure}[t] 
 \begin{center} 
 \includegraphics[width=\linewidth]{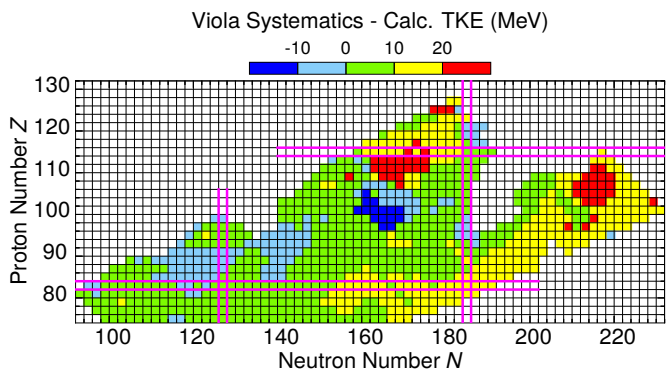} 
 \caption{Viola TKE systematics minus calculated average TKE.}
\label{vmctke} 
 \end{center} 
\end{figure}
\begin{figure}[b] 
 \begin{center} 
 \includegraphics[width=\linewidth]{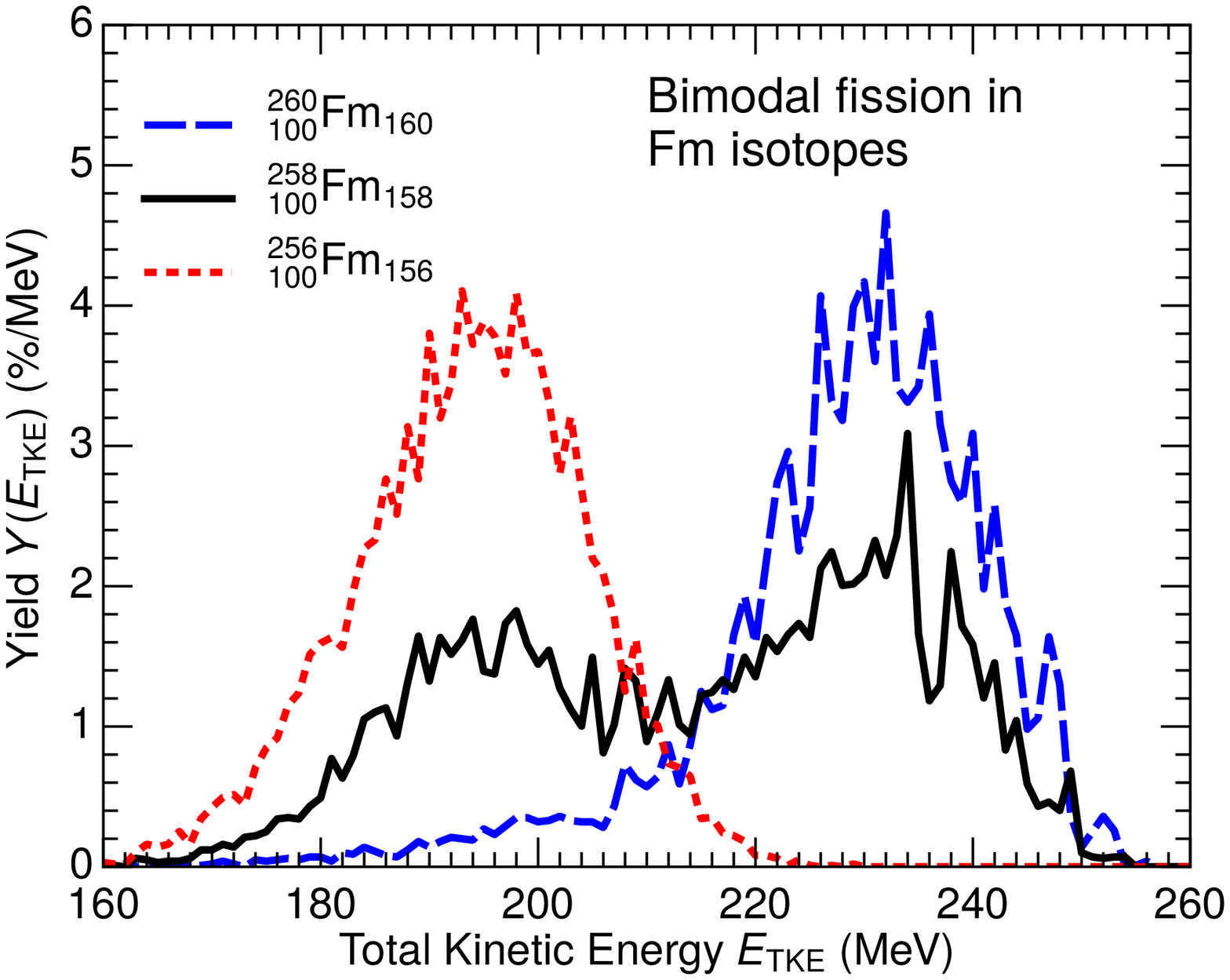} 
 \caption{Calculated fission-fragment total kinetic-energy distributions
   following low-energy fission
   of the three nuclides $^{256,258,260}$Fm.}
\label{yvstkefm}  
 \end{center} 
\end{figure}

\section{Calculated results}
\subsection{Fission-fragment mass distributions}
\label{sect:res_fragmass}

The calculated regions of symmetric and asymmetric fission in the full
region of study are shown in fig.\ \ref{s2pr}. The results agree very
well with fig.\ 3 in ref.\ \cite{moller15:b} in which the lower part
of the region in fig.\ \ref{s2pr} is shown. Figure \ref{s2pr} also shows
the upper part of the asymmetric-fissioning actinide region as well as
an additional region of asymmetry, approximately $108 \le Z \le 116$
and $164 \le N \le 176$. 
There are no experimental studies of the
whole region shown in fig.\ \ref{s2pr} but studies of 70 nuclides from
$Z=85$ to $Z=94$ were presented in ref.\ \cite{schmidt00:a}. 
It was suggested there that the transition between symmetric fission in the
lighter actinide region and asymmetric for heavier actinides is
$A\approx 226$. It is further stated that this is somewhat surprising
since one would expect both protons and neutrons to affect what
regions fission symmetrically or asymmetrically. 
However, in ref.\ \cite{schmidt00:a} fission mass distributions across the line
$A\approx 226$ are obtained for only a few proton numbers, namely $Z=89,90$ and 91.
Here, and in ref.\ \cite{moller15:b}, which covers a larger, contiguous region of nuclides than the experimental work, the results
show that both protons and neutrons affect asymmetry. 
Particularly interesting is that above $Z \approx 88$ ($N\approx 132$) the calculated transition
line is clearly not a constant mass number $A$, but approximately a
constant $N-Z$ for a range of about eight proton numbers.  This
prediction has yet to be tested experimentally. 

The calculated mass difference between the heavy and light fragments
is presented in fig.\ \ref{massdif}. Soon after the discovery of fission it was observed
that the mass of the heavy fragment remains relatively constant at $A\approx 140$,
as the mass number of the fissioning system evolves from $A\approx 230$ towards
heavier systems. Consequently the light mass increases so the heavy/light fragment
mass difference decreases as the fissioning system becomes heavier. 
This is particularly the case towards the neutron-deficient region where experimental data exists. 
Striking in fig.\ \ref{massdif} is the abrupt transition to a small region
(around $Z \approx 110, N \approx 166$)
of very large differences
between the 
heavy and light fragment masses as well as a very low 
symmetric to asymmetric yield ratio seen in fig.\ \ref{s2pr}.

This is illustrated in a complementary way in fig.\ \ref{fragmasses} where heavy and light fragment
mass pairs are plotted as coloured symbols. 
Figure \ref{fragmasses} shows that asymmetric fission of actinides with mass number from $A\approx220$ to $A\approx246$
corresponds to divisions into a $^{132}$Sn-like heavy fragment and the corresponding partner.
Similarly, there is a highly asymmetric region just below and at
$Z \approx 114$ corresponding to division into a $^{208}$Pb-like heavy fragment and the
corresponding partner.
In the superheavy region, a $^{208}$Pb-like fission fragment is compatible with recent results based on density functional theory 
\cite{warda18:a,matheson19:a},
whereas recent calculations using a prescission point model predicts divisions into a $^{132}$Sn-like light fragment and the corresponding partner \cite{carjan19:a}.

It was observed in ref.\ \cite{ichikawa05:a} that
the potential-energy surfaces for some superheavy nuclei, for example $^{272}$Ds
(see fig.\ 7 in ref.\ \cite{ichikawa05:a}), exhibit the usual fission valley but also
a fusion valley higher in energy than the fission valley and separated from
the fission valley by a pronounced ridge. Its existence might be
an additional reason that some ``cold'' fusion reactions have led
to evaporation residue creation; the ridge between fusion valley
and fission valley hinders the fusion trajectory to deflect into
the fission valley before a compound nucleus is created \cite{ichikawa05:a}.
It might be tempting to predict
symmetric fission because the fission valley for this nucleus corresponds to
symmetric shapes. On the other hand small neck radii are reached at much lower
$Q_2$ values in the fusion valley than in the fission valley. In this
and many other cases it is
obviously hard to predict fission mass divisions from our limited ability to
understand the details
of the calculated multi-dimensional potential-energy surface.
However, the BSM method constitutes a well-defined
approach to calculate fission mass yields. For many nuclei in this region
the result is highly asymmetric fission, paradoxically corresponding to
fission in the fusion valley. 

Some specific, calculated, yield distributions are presented
in figs.\ \ref{myields1} and \ref{myields2} in which the highly variable and
rapidly changing character of the distributions is readily apparent.
The calculated mass yields in fig.\ \ref{myields1} show several similarities with the yields obtained using a scission-point model \cite{carjan15:a,pasca18:a},
though the exact 
transition points from asymmetric to symmetric fission in this region differ slightly.
Nuclides around the superheavy region exhibit modes leading to both 
symmetric fission and highly asymmetric fission (cf. $(Z,N)=(108,162)$ in fig.\ \ref{myields2}).
Due to the subtle competition between these two modes,
there can be abrupt changes in yields between neighbouring nuclides ({\it e.g.} $(Z,N)=(108,156)$).

There are some observations of fission kinetic energies
in the superheavy region, normally in fission events that terminate
$\alpha$-decay chains, but it is hard to draw conclusions about the asymmetry of fission
due to the very few events observed. Some observations  of such fission events
are reported in 
ref.\ \cite{oganessian07:a}. 
Sometimes a high TKE value is associated
with symmetric fission, in particular for some Fm isotopes where the scission configuration
is very compact. However, if the fission configuration is ``liquid-drop-like'' symmetric
fission would correspond to elongated scission shapes and a low TKE. 
Therefore, a TKE value by itself
is not sufficient to establish if fission is symmetric or asymmetric. We show in fig.\ \ref{s2prshe}
an enlarged portion of fig.\ \ref{s2pr} and two examples of decay chains 
discussed in ref.\ \cite{oganessian07:a} and their termination by fission at the decay-chain endpoints.

Figure \ref{s2prfm} shows an enlarged display of the symmetric-yield to
peak-yield ratio in Fig. \ref{s2pr} to enhance details
in the vicinity of $^{258}_{100}{\rm Fm}_{158}$. 
The transition from asymmetric to symmetric fission 
at $^{258}_{100}{\rm Fm}_{158}$ is very visible in this figure as well as in fig.\ \ref{myields1}.
In this region there are also drastic variations of TKE distributions, so we now discuss the
results obtained for these distributions.

\subsection{Fission-fragment TKE distributions}
The calculated average TKE values (see sect.\ \ref{tkemeth} for details
about the method used) for the entire region of study are
shown in fig.\ \ref{tkeav}.
Obviously the TKE increases for heavier nuclides and towards
the neutron drip line. More easily interpretable is the difference
between the Viola TKE systematics \cite{viola85:a}
\begin{equation}
  \label{violatke}
    E_{\rm TKE}^{\rm (Viola)}=0.1189 Z^2/A^{1/3}+7.3 \text{ MeV},
    \end{equation}
and the actually calculated average TKE.
This is illustrated in fig.\ \ref{vmctke}. 
The substantial abrupt local variations seen in the region near $^{258}$Fm
are also seen experimentally \cite{john71:a,hoffman80:a,hulet80:a,hulet86:a,hulet89:a}.
In their main features, the measurements are well reproduced in the calculations.
In the region of very asymmetric fission below $Z=114$ ($N\approx$ 162--174) the TKE is lower than the
systematics as would be expected. Also in the heavy neutron-rich region ($N\approx$ 210--220), where
we obtain asymmetric fission yields, the average TKE is lower than given by the Viola
systematics.

Figure \ref{yvstkefm} shows
in detail the TKE distributions for the three
isotopes $^{256}$Fm, $^{258}$Fm, and $^{260}$Fm. The transition from ``normal'' TKE to
high TKE is dramatic, with $^{258}$Fm exhibiting a clear ``bimodal'' structure,
as has been seen experimentally \cite{hulet86:a}.
A bimodal structure in the Fm region has also been obtained within the scission-point model \cite{carjan15:a}.

\section{Summary}
The BSM method \cite{randrup11:a}, which in systematic calculations
was previously applied to nuclides from $Z=74$ to $Z=94$ \cite{moller15:b},
has here been used to perform systematic yield calculations 
in the region $74 \leq Z \leq 126$ and $92 \leq N \leq 230$
for 896 even-even nuclides.
Where there is overlap with the previous calculations there is good agreement.
Results above $Z=94$ show for the first time predictions based on
the BSM method for the heavier actinide region and the superheavy region.
In the vicinity of $Z=114$ there is a new, smaller region of asymmetry corresponding to $^{208}$Pb-like heavy fragments.
Neutron-rich nuclei near the neutron  drip line are also 
predicted to fission asymmetrically.

\begin{acknowledgement}
The authors are grateful to D. Rudolph and C. Schmitt for valuable discussions and suggestions.
This work was supported by the Swedish Natural Science Research Council (S.{\AA}.) and the Knut and Alice
Wallenberg Foundation (grant No. KAW 2015.0021) (M.A., B.G.C. and S.{\AA}.); J.R. was supported in part by the NNSA DNN R\&D of the U.S. Department of Energy
and acknowledges a Visiting Professorship at Mathematical physics at Lund University.
\end{acknowledgement}

 \bibliographystyle{unsrt}
 \bibliography{refer-crank.bib}
\end{document}